\documentclass[preprint,review, 12pt]{elsarticle}
\usepackage{graphicx}
\usepackage{amssymb}
\usepackage{amsthm}
\usepackage{amsmath}

\journal{Physica B}

\begin{document}

\begin{frontmatter}


\title{Collective  resonances in plasmonic  crystals: Size matters}
%% \author{M. C. Schaafsma\corref{cor1}\fnref{label2}}

  \author{S. R. K. Rodriguez $^\dag$\footnote{$^\dag$These authors had an equal contribution}}
  \address{Center for Nanophotonics, FOM Institute AMOLF, c/o Philips Research Laboratories, High Tech Campus 4, 5656 AE Eindhoven, The Netherlands.}

  \author{M. C. Schaafsma $^\dag$}
  \address{Center for Nanophotonics, FOM Institute AMOLF, c/o Philips Research Laboratories, High Tech Campus 4, 5656 AE Eindhoven, The Netherlands.}

  \author{A. Berrier}
  \address{Center for Nanophotonics, FOM Institute AMOLF, c/o Philips Research Laboratories, High Tech Campus 4, 5656 AE Eindhoven, The Netherlands.}

  \author{J. G\'{o}mez Rivas}
  \address{Center for Nanophotonics, FOM Institute AMOLF, c/o Philips Research Laboratories, High Tech Campus 4, 5656 AE Eindhoven, The Netherlands}
   \address{COBRA Research Institute, Eindhoven University of Technology, P.O. Box 513, 5600 MB Eindhoven, The Netherlands}
  \ead{rivas@amolf.nl}
    \date{\today}

\begin{abstract}
Periodic arrays of metallic  nanoparticles may sustain  Surface
Lattice Resonances (SLRs), which are collective resonances
associated with the diffractive coupling of Localized Surface
Plasmons Resonances (LSPRs). By investigating a series of arrays
with varying number of particles, we traced the evolution of SLRs to
its origins. Polarization resolved extinction spectra of arrays
formed by a few nanoparticles were measured,  and found to be in
very good agreement with calculations based on a coupled dipole
model. Finite size effects on the optical properties of the arrays
are observed, and our results provide insight into the
characteristic length scales for collective plasmonic effects: for
arrays smaller than $\sim5\times5$ particles, the $Q$-factors of
SLRs are lower than those of LSPRs; for arrays larger than
$\sim20\times20$ particles, the $Q$-factors of SLRs saturate at a
much larger value than those of LSPRs; in between, the $Q$-factors
of SLRs are an increasing function of the number of particles in the
array.
\end{abstract}


\end{frontmatter}

% \linenumbers

\section{Introduction}

Sophisticated methods for manipulating light at the nanoscale are
increasingly developed in the field of metallic
nano-optics~\cite{Soukoulis11}. Although metals may provide
advantages over dielectrics associated with the large
electromagnetic enhancements they may create~\cite{Aizpurua}, the
high losses accompanying resonant effects pose a serious challenge
for their emergence as a viable technology~\cite{Soukoulis2010}. The
development of resonances with high quality factor $Q$ is therefore
of great relevance in the field of nanoplasmonics. One way to
minimize losses in plasmonic systems is based on collective
resonances~\cite{Ropers}, which leads to a modification of radiative
damping - the dominant contribution to the plasmon linewidth. In the
case of periodic arrays of metallic nanoparticles, it was calculated
that near the critical energy where a diffraction order changes from
radiating to evanescent in character, dipolar interactions would
lead to the emergence of a new, narrow linewidth plasmonic
resonance~\cite{Meier85}. Carron $et$ $al.$ seem to have been the
first to investigate this phenomenon experimentally~\cite{Carron},
but the resonances were not as sharp as predicted by the theory due
to technological limitations rendering imperfect structures. Schatz,
Zou, and co-workers revived the interest in these lattice-induced
plasmonic resonances with a series of theoretical papers mainly
based on the Coupled Dipole Model (CDM). Extinction efficiencies
higher than 30 were predicted~\cite{Zou04}, but experimental
observation of these narrow resonances remained
elusive~\cite{Kall05}. These resonances are now known as Surface
Lattice Resonances (SLRs), and they have been observed
experimentally in the recent years by several
groups~\cite{Barnes08,Crozier08,Kravets08,Vecchi09}. SLRs arise from
the diffractive coupling of Localized Surface Plasmon Resonances
(LSPRs) of individual particles. This coupling is mediated by
Rayleigh anomalies, which correspond to the condition whereby a
diffracted wave propagates in the plane of the array. The properties
of SLRs, just as those of LSPRs, generally depend on size, geometry
and composition of the particle, and on the surrounding medium and
polarization of the light field~\cite{Barnes08}. Moreover, due to
their collective nature, SLRs rely strongly on the interparticle
distance and on the long-range order in the lattice~\cite{Barnes09}.
A question that remains open is how the number of particles in the
array influences the properties of SLRs.  Remarkable insight into
this problem has been obtained for the complementary structures of
subwavelength hole arrays in metallic films ~\cite{Bravo06,
Ebbesen08}. However, the influence of finite size effects on the
optical properties of nanoparticle arrays sustaining collective
resonances has not been discussed yet. Although a similar response
is expected for nanohole and nanoparticle arrays based on Babinet's
principle~\cite{GarciadeAbajo10}, we highlight that there is a
fundamental difference between the two systems. Namely, whereas
radiative coupling in nanohole arrays may take place via surface
plasmon polaritons propagating through continuous metallic films,
nanoparticle arrays consist of isolated metallic islands that may
electromagnetically couple through diffraction.

In this paper, we present an experimental and theoretical study on
the evolution of SLRs as a function of the number of particles in
the array. We investigate finite size effects on the extinction
spectra starting at the smallest array size, i.e., 2x2 particles.
This work is organized as follows. In section 2 we describe the
samples and experimental methods. In section 3 we provide an
overview of the Coupled Dipole Model (CDM), which we use to
calculate the extinction spectra of various arrays. In section 4 we
compare the measurements to the CDM calculations.

%It should be noted that owing to the weak signals produced in
%extinction by small nanoparticle arrays, studies on small arrays
%have focused on Rayleigh scattering spectra\cite{Kall05}.
%Transmittance measurements have been mainly obtained for large
%arrays, which we may call `infinite', in as much as further addition
%of particles to the array does not contribute in any significant
%manner to the extinction efficiency. Nevertheless, because
%$extinction = scattering + absorption$ is a more complete
%description of the system's light matter interactions~\cite{Bohren},
%transmittance spectra from which extinction may be directly obtained
%are of significant importance. Herein, we present such spectra, and
%discuss the critical length scales over which finite size effects on
%2D arrays of nanoparticles are important.

%When particles are in close proximity to each other, near field SP coupling may lead to giant electromagnetic enhancements, or so-called hot-spots. These hot-spots are ideal for enhancing the emission of molecules, SERS, etc.
%Although the field enhancements can reach very large values, this effect usually occrus within a small volume.

\section{Sample and Experimental Methods} \label{sample}
A series of gold nanodisk arrays with dimensions of  $N \times N$
particles, $N$ ranging from 2 to 10, were fabricated by electron
beam lithography onto an amorphous quartz substrate. The nanodisks
have a height of 50 nm, a diameter of 120 nm, and they are arranged
in rectangular arrays with lattice constants $a_x = 500$ nm and $a_y
= 300$ nm. In order to surround the arrays by a homogeneous medium,
we evaporated 150 nm of silica on top, added index matching fluid of
$n= 1.45$, and placed an amorphous quartz superstrate identical to
the substrate. The homogeneous environment was created in order to
enhance the diffractive coupling of the
particles~\cite{GarciadeAbajo10}.

A confocal microscope with a two-axis translation stage with a
precision of 200 nm was used for the extinction experiments. A white
light beam from a halogen lamp impinged onto the samples at normal
incidence, rendering a plane wave excitation. The transmitted light
was collected by the microscope's objective and directed into a
fiber coupled spectrophotometer. The polarization of the incident
light was set parallel to the $a_y = 300$ nm pitch, so that
diffractive coupling along the $a_x = 500$ nm may take place. In
order to detect the full extinction of the incident plane wave, the
angle subtended by the detector should be sufficiently
small~\cite{Bohren}. This translates into the requirement for a
small Numerical Aperture (NA) and a small aperture of the confocal
pinhole in the collection path, which poses a challenge for
detecting the low signals produced in extinction by small arrays. We
used a $20\times$, NA=0.4 objective, and a $60\mu$m confocal
pinhole, which represents an improvement relative to studies with a
higher NA's~\cite{Crozier08}. The Field of View (FoV) seen by the
detector was determined by the knife-edge technique, in which the
transmitted power is measured across the boundary between a
transparent surface and a metallic layer; we found a $\textrm{FoV} =
6.0 \pm 0.5$ $\mu m$.

\section{Theory}\label{theory}
The Coupled Dipole Method (CDM)  under the Modified Long Wavelength
Approximation (MLWA) is widely used for calculating extinction
spectra of nanoparticle arrays~\cite{Jensen99,Zou04}. The CDM is
based on modeling each particle as a radiating dipole, and
calculating its interaction with the fields radiated by all other
dipoles in the array. We herein provide an overview of the CDM.

\subsection{Polarizability of a single particle}
Since no closed-form solution exists for the polarizability of
cylinders, the nanodisks are approximated as oblate
spheroids~\cite{Venermo}. The static polarizability of an ellipsoid
along one of its main axes is given by
\begin{equation}
\alpha^{static}=\frac{(\epsilon_{p}-\epsilon_{d})}{[3(\epsilon_{p}-\epsilon_{d})\cdot
L+3\epsilon_{d}]}\cdot V \;,
\end{equation}
with $\epsilon_{p}$ and $\epsilon_{d}$ the relative dielectric
constants of the particle and of the surrounding dielectric,
respectively, $V$ the volume of the particle, and $L$ a form factor
for each of the three main axes of the ellipsoid~\cite{Hulst}. In
the MLWA, the static polarizability is  modified to account for
dynamic depolarization and radiative damping. This polarizability is
given by
\begin{equation}
\alpha=\frac{\alpha^{static}}{1-\frac{2}{3}ik^3\alpha^{static} -
\frac{k^2}{r}\alpha^{static}} \;, \label{mlwa}
\end{equation}
where $k=2\pi n/\lambda$ is the incident wave vector with $\lambda$
the vacuum wavelength and $n$ the refractive index of the
surrounding dielectric, and $r$ is the radius of the main axis of
the ellipsoid in the direction of the polarization. The term
$\frac{2}{3}ik^3\alpha^{static}$ corresponds to dynamic
depolarization, and the term $\frac{k^2}{r}\alpha^{static}$ accounts
for radiative damping~\cite{Jensen99}.

\subsection{Coupled Dipole Model (CDM)}
From the polarizability $\alpha_{i}$ at position $r_i$ of the array,
the polarization of the medium $P_i$ can be calculated with the
relation $P_{i}=\alpha_{i}E_{loc,i}$, where $E_{loc,i}$ is the local
field, given by the sum of the incident field and the fields
scattered by all other particles. Under plane wave illumination, the
local field can be expressed as
\begin{equation}
 E_{loc,i}=E_{0} e^{ikr_i} - \sum_{\substack{j\neq i \\ j=1}}^{N^2} A_{ij}\cdot
 P_{j} \;,
\end{equation}
where $A_{ij}$ represents the dipolar interaction between particles
$i$ and $j$, and its dot product with $P_j$ reads,
\begin{align}\label{Aij}
A_{ij}\cdot P_{j} &= k^2 e^{ikr_{ij}}\frac{r_{ij}\times(r_{ij}\times P_{j})}{r_{ij}^3} \nonumber \\
 & + e^{ikr_{ij}}(1-ikr_{ij})\frac{r_{ij}^3P_{j}-3r_{ij}(r_{ij}\cdot P_{j})}{r_{ij}^5}, \\
  & (i,j = 1 \ldots  N^2, i \neq j)\;, \nonumber
\end{align}
for an $N \times N$ particle array. The polarization vector
$\tilde{P}$ can be calculated from,
\begin{equation}\label{eq:general}
 \tilde{E}= \tilde{\tilde{A'}} \tilde{P} \;,
\end{equation}
where the diagonal terms of $\tilde{\tilde{A'}}$ are
$A_{ii}=\alpha_{i}^{-1}$, and the off-diagonal terms of
$\tilde{\tilde{A'}}$ are calculated from Equation~\ref{Aij}. From
the polarization vectors, the extinction cross section of the array
can be calculated using
\begin{equation}\label{Cext}
 C_{ext} = \frac{4\pi k}{|E_{0}|^2}\sum_{i=1}^{N^2}Im(E^{*}_{inc,i}\cdot P_{i}) \;,
\end{equation}
where $E^{*}_{inc,i}$ is the complex conjugate of the incident
field~\cite{Bohren}.  An extinction efficiency $\eta_{ext}$ can be
obtained by normalizing the extinction of the array to the sum of
the geometrical cross section of the particles, i.e.,
\begin{equation}
\eta_{ext}= \frac{C_{ext}}{N^2 A_p} \;,
\end{equation}
with $A_p$ the cross sectional area of each particle.

In order to compare theoretical extinction spectra with experimental
transmittance spectra, we define the $Extinction = 1-T$, with $T$
the transmittance in the forward direction within the NA of the
objective. To obtain a transmittance from Equation~\ref{Cext} that
may result in quantitative agreement with experimental data, the
$FoV$ in the experiments needs to be considered. In our case, the
$FoV$ is equal for all arrays, and larger than the largest array.
This leads to a transmittance given by
\begin{equation}
\label{T}
 T = 1 - \frac{C_{ext}}{FoV} \;.
\end{equation}

    Although the focus of the present work is on finite size effects in
nanoparticle arrays, it is instructive to assess how well the
response of very large, but finite, arrays converges to the response
of an `infinite' array. For `infinite' arrays illuminated at normal
incidence, Eq.~\ref{eq:general} can be simplified  assuming that the
induced polarization in all particles is equal. This leads to an
effective polarizability $ \alpha^{\star}$, given by~\cite{Zou04}
\begin{equation} \label{alpha_eff}
\alpha^{\star} =  \frac{1}{1/ \alpha - S} \;,
\end{equation}
where $S$ is the retarded dipole sum given by
\begin{equation} \label{S}
S = \sum_{j \neq i}  \frac{ (1 - i k r_{ij} )(3 \textrm{cos}^2
\theta_{ij} - 1 ) e^{i k r_{ij} } }{r_{ij}^3} + \frac{ k^2
\textrm{sin}^2 \theta_{ij} e^{i k r_{ij} } }{r_{ij}} \;,
\end{equation}
with $ \theta_{ij}$ the angle between $r_{ij}$ and the polarization
direction. Note that $S$ is a purely geometrical factor through
which the lattice modifies the polarizabilities of the individual
particles, so that when $1/ \alpha = S$ a pole in $ \alpha^{\star}$
occurs, giving rise to a lattice-induced resonance. Because $\alpha$
and $S$ are in general complex numbers, strict equality does not
occur and $ \alpha^{\star}$  remains finite. However, a resonance
indeed arises when the real part of  $1/ \alpha -  S$ vanishes.

\section{Results}
\begin{figure}
\centerline{\includegraphics[width=8cm]{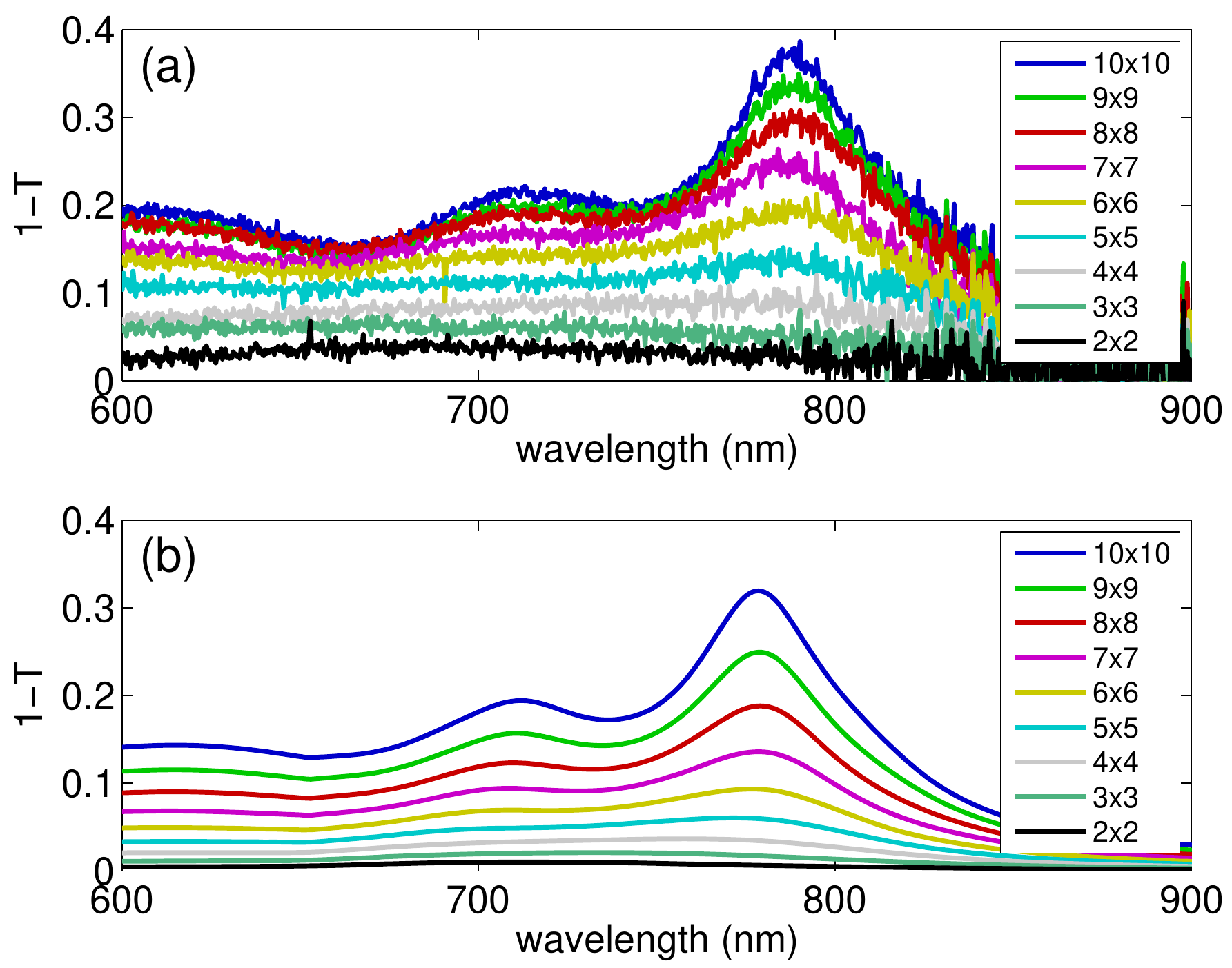}}
\caption{(a)Measurements, (b) Coupled Dipole Model calculations, of
the extinction spectra of $N$x$N$ gold nanodisk arrays, where $N$
ranges from 2 to 10. } \label{fig:meas}
\end{figure}
Figure 1 shows extinction spectra as obtained from measurements (a),
and CDM calculations (b), for the arrays described in
section~\ref{sample}. At $\lambda \approx 780$ nm we observe a peak
gradually growing in extinction as the array size increases. This
peak is the SLR, becoming increasingly sharper due to a collective
suppression of radiative damping. The local minimum seen in
extinction at $\lambda \approx 745$ nm corresponds to the Rayleigh
anomaly. We also note a secondary maximum in extinction observed at
$\lambda \approx 720$ nm. F\'{e}lidj $et$ $al$. observed a similar
feature which, together with the primary resonance, they attributed
to two diffracted waves into different media~\cite{Felidj05}.
However, the observation of a very similar feature by  Augui\'{e}
and Barnes in a fully homogeneous medium~\cite{Barnes08} questioned
the explanation provided by F\'{e}lidj $et$ $al$. At this point, we
would like to clarify that to obtain the good agreement  between
measurements and calculations seen in Figure 1, we did not use
normally incident illumination alone. Instead, the spectra in  Fig.
1(b) were calculated by averaging the extinction over angles of
incidence between $\theta= 0^{\circ}$ and $6^{\circ}$. This implies
that the incident beam in the experiment was not well collimated.
The motivation to angularly average the extinction spectra, and its
connection with the secondary minimum seen in Refs.~\cite{Barnes08,
Felidj05}, is discussed next.

Figure 2(a) shows the calculated variable angle extinction
efficiency spectra for an array of 10x10 particles.  Two sharp
resonances are seen in the spectra, displaying an anti-crossing
behavior near normal incidence associated with their mutual
coupling. The peak in extinction  at $\lambda \approx 780$ nm and
$\theta= 0^{\circ}$ is the SLR associated with the (-1,0)
diffraction order - this is the primary peak in the spectra of Fig
1(b). The peak in extinction at $\lambda \approx 700$ nm  and
$\theta= 6^{\circ}$ is the  (+1,0) SLR - this is secondary peak in
extinction. Thus, we observe that the coupling between the (+1,0)
and (-1,0) SLRs leads to the opening of a stop-gap in the dispersion
relation of the array. The mutual coupling between bright and dark
SLRs (in `infinite' arrays) and the consequent opening of the gap
were recently discussed~\cite{Rodriguez11}. By bright/dark it is
meant that the resonance couples efficiently/inefficiently to light,
which results from a symmetric/antisymmetric field distribution.
Notice that only for angles of incidence of ~$\theta \gtrsim
2^{\circ}$, the (-1,0) SLR center wavelength increases with the
angle. The flattening of the band observed near normal incidence
indicates a reduction of the mode's group velocity and the formation
of standing waves. This is the origin of the high extinction from
the (-1,0) SLR at normal incidence, since the density of optical
states is enhanced at the band-edge. On the other hand,  the (+1,0)
SLR can not be excited by a normally incident plane wave; a
narrowing of the plasmon linewidth and a diminishing extinction are
seen as the angle of incidence appproaches $\theta= 0^{\circ}$. The
observed behavior is characteristic of subradiant damping, whereby
radiative damping is suppressed due to antisymmetric field
distributions~\cite{Ropers}. In Fig. 2(b) we display cuts in angle
at $\theta =0^{\circ}$, $\theta=6^{\circ}$ of Fig. 2(a), and average
the spectra between $\theta= 0^{\circ}$ and $\theta=6^{\circ}$. By
comparing the three curves in Fig. 2(b) with the experimental data
in Fig. 1(a) for the 10x10 array, it is clear that the best
agreement between measurements and calculations is obtained by
assuming that the incident beam contains a distribution of
k-vectors, which on a best-fit basis, we estimated to include angles
of up to $6^{\circ}$. Therefore, the variable angle extinction
spectra in Fig. 2 explain the features we observe in the
experiments, and also elucidate what is most likely the origin of
the secondary peak in extinction seen in Refs.~\cite{Barnes08,
Felidj05}, and also in other studies~\cite{Kravets08},i.e., an
non-perfectly collimated incident beam.
\begin{figure}
\centerline{\includegraphics[width=8cm]{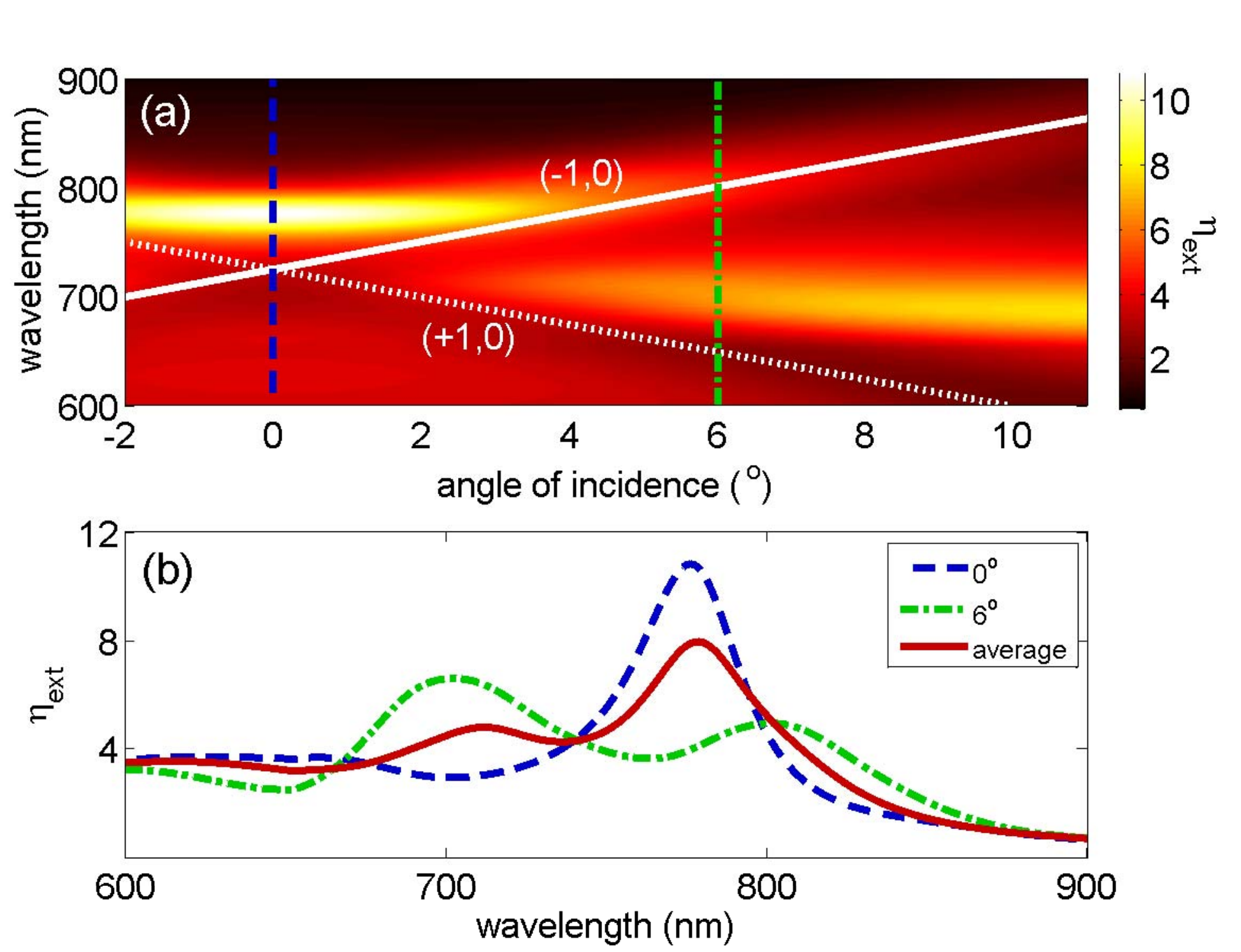}} \caption{
Coupled Dipole Model calculations for an array of 10x10 gold
nanodisks. (a) Variable angle extinction spectra, (b) the dashed and
dashed-dotted curves are cuts at $\theta = 0^{\circ}$ and $\theta =
6^{\circ}$ of (a), respectively, and the solid curve is the average
extinction between 0 and 6 degrees.} \label{fig:dispersion}
\end{figure}

As observed in Fig. 2 and discussed in Ref.~\cite{Rodriguez11}, SLRs
are Fano resonances~\cite{Fano}, i.e., asymmetric resonances molded
by the interference between two channels: the scattered intensity in
the plane of the array at the Rayleigh anomaly condition, and the
background transmission. Although Fano described quantum
interference phenomena, his model has found broad applicability to
classical systems also, and particularly in
plasmonics~\cite{Genet03, Fano10}. We have fitted the extinction
efficiency spectra for each array with a Fano equation of the form
\begin{equation}
\eta_{eff} = C_0 \frac{ [q + 2(\omega - \omega_0)/\Gamma]^2 }{ 1 +
[2(\omega - \omega_0)/ \Gamma]^2},
\end{equation}
with $C_0$ an amplitude constant, $\omega_0$ the resonance center
frequency,  and $\Gamma$ the linewidth~\cite{Galli09}. From the
fitted values, we have calculated a quality factor
 $Q_{SLR} = \omega_{0} / \Gamma $, which is shown in Fig. 3(a) as a function of the
number of particles $N$ along each dimension of the array. This
expression for the quality factor defines the ratio of the stored to
dissipated energy only in Lorentzian resonators. Nevertheless, we
use it to characterize the spectral narrowing and to obtain a
qualitative insight into the suppression of radiative damping as the
size of the array increases. For comparison, we plot as horizontal
lines in Fig. 3(a) the calculated $Q$ of a single particle, $Q_s$,
and  of an `infinite' array, $Q_{\infty}$. In Fig. 3(b) we show the
extinction spectra for some of the finite arrays to illustrate how
the SLR lineshapes evolve as $N$ increases. In Fig. 3(c) we show the
single particle and `infinite' array spectra; the latter was
calculated using Eqs.~\ref{alpha_eff} and~\ref{S}. Figure 3 shows
how the increasing extinction at the SLR wavelength and the
simultaneous spectral narrowing lead to an increasing $Q_{SLR}$.

%It should be noted that this expression for the $Q$ has its origin
%in the nature of the Lorentzian function, whose Fourier transform
%gives an exponentially decaying function with a characteristic time
%$\tau$. For a Lorentzian resonance, the $Q$ is therefore also
%related to the lifetime of the excited state, $\tau$, according to
%the relation $Q = 2 \pi \nu \tau$. The equivalence between these two
%expressions is transparent for a Lorentzian Resonance - both
%rendering the ratio of stored to dissipated energy - but not for a
%Fano resonance. Nevertheless, qualitative agreement is
%unquestionable, so our definition $Q = \omega_{res} / \Gamma $ with
%$\Gamma$ the linewidth extracted from Fano's model should be taken
%not as a quantitative result, but as a means to obtain qualitative
%insight into the suppression of radiative damping as the array size
%increases.

\begin{figure}
\centerline{\includegraphics[width=8cm]{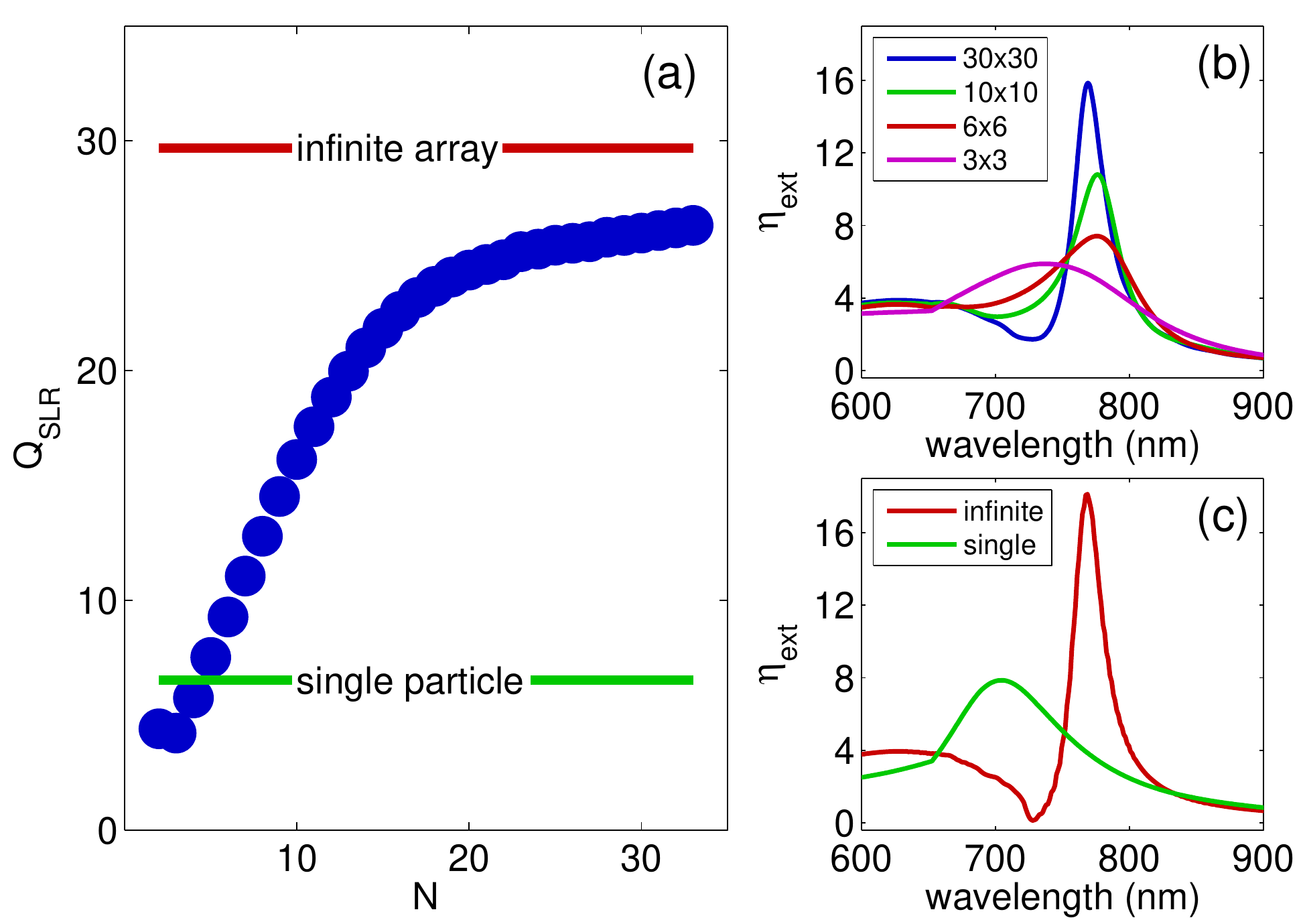}}
\caption{(a) Calculated quality factor of SLRs ($Q_{SLR}$) as a
function of the number of particles ($N$) along each dimension of
the array. (b) Calculated extinction spectra for some of the arrays
yielding the increasing $Q_{SLR}$ in (a). (c) Calculated extinction
spectra of a single particle and an infinite array. The quality
factors of the single particle and the infinite array are indicated
by the horizontal lines in (a).} \label{fig:Q}
\end{figure}

From the values reported in Fig. 3(a), we may identify three regimes
for $Q_{SLR}$: i) For $N<5$, $Q_{SLR} < Q_s$; ii) For $5<N<20$,
$Q_{SLR}$ increases rapidly; And iii) for $N>20$, $Q_{SLR}$ begins
to saturate, slowly approaching the value $Q_{\infty} \approx 30$ of
the `infinite' array. We understand the evolution of $Q_{SLR}$
through the three regimes as follows. In the first regime,
collective effects are weak. The extinction displays several peaks
of similar magnitude which together yield a `linewidth'
significantly broader than that of the single particle resonance.
The weak collective behavior is due to the large fraction of
particles located at the edges of the array. Although this regime
was neither measured nor identified in Ref.~\cite{Fedotov08}, the
authors discussed the underlying mechanism by which a decreasing
array size leads to a reduction in $Q$ for a collective resonance:
scattering losses at the edges of the array. In the second regime,
the fraction of particles at the edges decreases and the number of
particles resonating collectively increases, which lead to a rapid
increase in $Q_{SLR}$. In the third regime, further addition of
particles to the array does not affect drastically $Q_{SLR}$.
Although the in plane scattering losses continue to diminish, the
out-of plane scattering losses and material losses become the
dominant contribution to the linewidth. This can be intuitively
understood by considering the propagation lengths for the surface
polaritons associated with SLRs. In Ref.~\cite{Vecchi09b},
propagation lengths on the order of $\sim 10$ $\mu$m were found for
similar arrays of gold nanoparticles. In the present work,
diffractive coupling takes place along the $a_x = 500$ nm pitch of
the structure, which means that $10$ $\mu$m corresponds to 20 unit
cells. Not coincidentally, $Q_{SLR}$ begins to saturate near $N=20$.
Since further addition of particles does not contribute
significantly to the extinction, the arrays become effectively
`infinite'.

Figure 4 shows results from Finite Element Method (FEM) simulations
(COMSOL) for an `infinite' array of nanodisks as those in the
measurements and CDM calculations. Periodic boundary conditions were
used, and the particles are illuminated by a normally incident plane
wave. The transmittance spectra was first calculated, and it was
found to be in excellent agreement with the `infinite' array spectra
obtained from the CDM (Fig. 3(c)). In Fig. 4, we present the field
enhancement (in color scale) and the scattered field (arrows) at a
plane intersecting the particles at their midheight, for two values
of the vacuum wavelength. Figure 4(a) shows the spectra at $\lambda
= 670$ nm, which corresponds to the broad and weak feature seen in
extinction. As it can be recognized from the field pattern, a
dipolar resonance is excited in the individual particles. A small
field enhancement is observed near the surface of the particles, but
the region in between the particles exhibits field suppression. The
particles are therefore individually resonant with the incident
field and there is no collective behavior. This is the typical
behavior of LSPRs. In Fig. 4(b) we show the fields at $\lambda =
780$ nm, which corresponds to the SLR wavelength. Notice that a
dipolar resonance is excited as well, having an almost identical
radiation pattern to that in Fig. 4(a), but with two important
differences: The field enhancements are much larger (almost by an
order of magnitude), and the scattered intensity in the area between
the particles is much higher.  This has important consequences for
applications in sensing and modified spontaneous emission where
molecules may profit from a high density of states over extended
volumes.

In conclusion, we have demonstrated finite size effects in the
optical properties of metallic nanoparticle arrays. The critical
length scales over which collective effects are important were
discussed. We have identified an array size-dependent quality factor
for surface lattice resonances arising from the diffractive coupling
of localized surface plasmons ($Q_{SLR}$). $Q_{SLR}$ is lower than
the $Q$ of the single particle resonance for arrays smaller than 5x5
particles. For arrays sizes between 5x5 and 20x20 particles,
$Q_{SLR}$ dramatically increases beyond the value of $Q_s$,
saturating for larger arrays.

\begin{figure}
\centerline{\includegraphics[width=9.5cm]{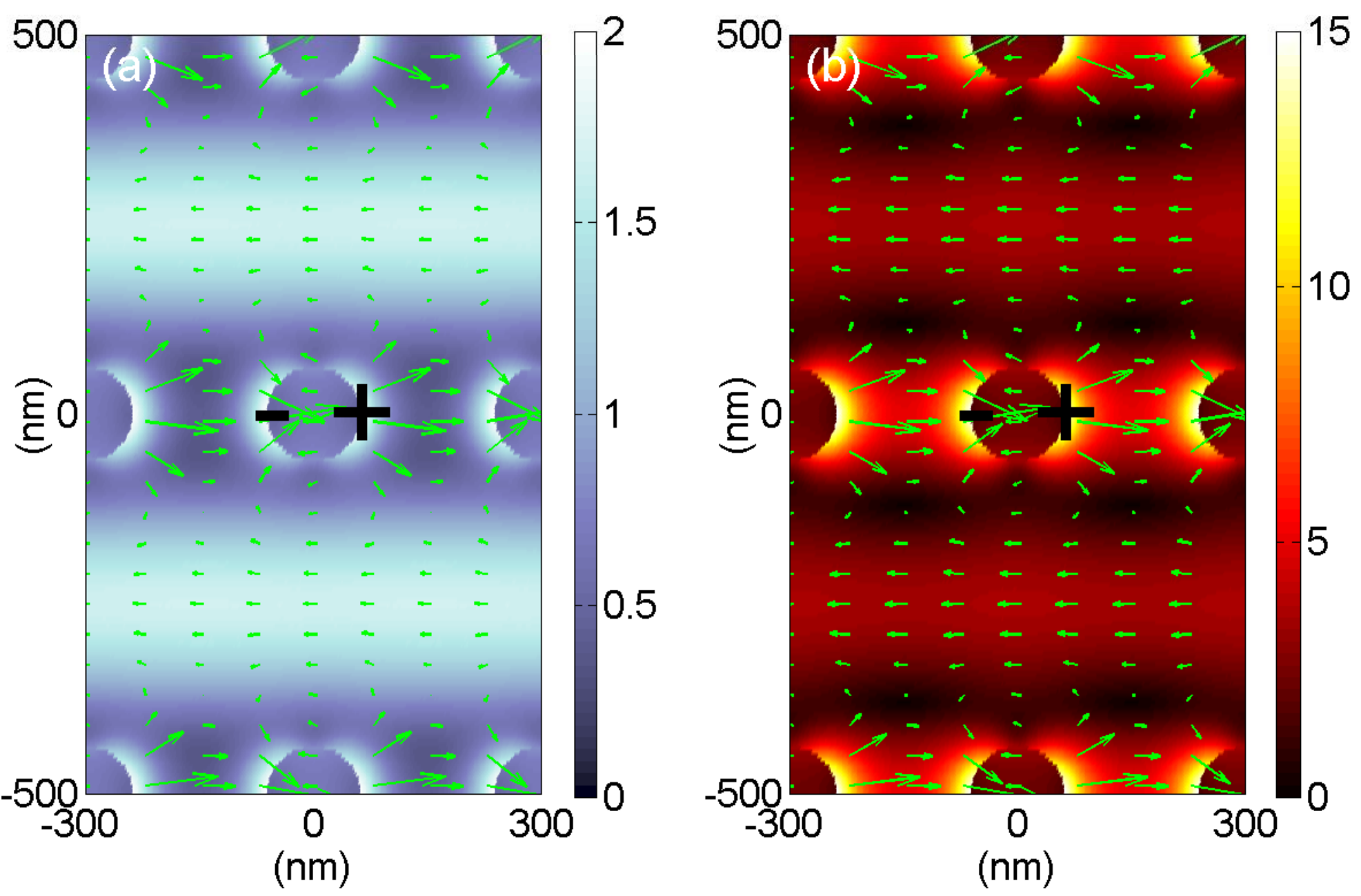}} \caption{
Total field enhancement (color scales) and scattered field (arrows)
for an infinite array of gold nanodisks surrounded by amorphous
quartz. Both plots are at a plane intersecting the nanodisks at
their midheight. The illumination is a normally incident plane wave
with a vacuum wavelength of (a)$\lambda = 670$ nm, and (b) $\lambda
= 780$ nm, which correspond to the wavelengths of the LSPR and SLR,
respectively. The polarization vector is along the same axis as the
excited dipole moment in the individual particles, indicated by the
$+$ and $-$ signs. } \label{fig:comsol}
\end{figure}

\section{Acknowledgements}
We would like to thank Costas Soukoulis for his truly inspiring
scientific works and conferences that have had a deep impact in a
whole generation of nanophotonics researchers. We would like to
acknowledge Bas Ketelaars for assistance in the fabrication of the
samples. This work was supported by the Netherlands Foundation
Fundamenteel Onderzoek der Materie (FOM) and the Nederlandse
Organisatie voor Wetenschappelijk Onderzoek (NWO), and is part of an
industrial partnership program between Philips and FOM.
\newpage
\bibliographystyle{elsarticle-num}
%\bibliographystyle{model1a-num-names}
%\bibliography{finite4}

\end{document}